\documentclass[%
reprint,
superscriptaddress,
frontmatterverbose,
amssymb,amsfonts,amsmath,
footinbib,
aip,
nofootinbib
]{revtex4-1}

\usepackage{graphicx}
\usepackage{dcolumn}
\usepackage{bm}
\usepackage{epsfig}
\usepackage{epstopdf}
\usepackage{natbib}
\usepackage{xfrac}
\usepackage[version=4]{mhchem}
\usepackage{soul}
\usepackage[mathlines]{lineno}
\usepackage{color,soul}
\usepackage[hidelinks=true]{hyperref}
\hypersetup{colorlinks   = true, urlcolor     = blue, linkcolor    = blue,  citecolor   = red }

\newcommand{\Var}{\mathrm{Var}}
\newcommand{\bra}[1]{\left(#1\right)}
\newcommand{\avr}[1]{\left<#1\right>}
\newcommand{\Bra}[1]{\left[#1\right]}

\begin{document}
	
	
	\title{Spatial heterogeneity may form an inverse camel shape Arnol'd tongue in parametrically forced oscillations}
	
	\author{Yuval Edri} \altaffiliation[Current address: ]{Laboratory of Sensory Neuroscience, The Rockefeller University, New York, NY 10065, USA}
	\affiliation{Department of Solar Energy and Environmental Physics, Blaustein Institutes for Desert Research (BIDR), Ben-Gurion University of the Negev, Sede Boqer Campus, Midreshet Ben-Gurion, 8499000, Israel}
	\author{Ehud Meron}%
	\affiliation{Department of Solar Energy and Environmental Physics, Blaustein Institutes for Desert Research (BIDR), Ben-Gurion University of the Negev, Sede Boqer Campus, Midreshet Ben-Gurion 8499000, Israel}
	\affiliation{Department of Physics, Ben-Gurion University of the Negev, Be'er Sheva 8410501, Israel}
	\author{Arik Yochelis}%
	\email{yochelis@bgu.ac.il}
	\affiliation{Department of Solar Energy and Environmental Physics, Blaustein Institutes for Desert Research (BIDR), Ben-Gurion University of the Negev, Sede Boqer Campus, Midreshet Ben-Gurion 8499000, Israel}
	\affiliation{Department of Physics, Ben-Gurion University of the Negev, Be'er Sheva 8410501, Israel}

	\date{\today}
	
	\begin{abstract}
		Frequency locking in forced oscillatory systems typically occurs in `V’-shaped domains in the plane spanned by the forcing frequency and amplitude, the so-called Arnol'd tongues. Here, we show that if the medium is spatially extended and monotonically heterogeneous, e.g., through spatially-dependent natural frequency, the resonance tongues can also display `U' and `W' shapes; to the latter, we refer as ``inverse camel'' shape. We study the generic forced complex Ginzburg--Landau equation for damped oscillations under parametric forcing and, using linear stability analysis and numerical simulations, uncover the mechanisms that lead to these distinct shapes. Additionally, we study the effects of discretization, by exploring frequency locking of oscillators chains. Since we study a normal-form equation, the results are model-independent near the onset of oscillations, and, therefore, applicable to inherently heterogeneous systems in general, such as the cochlea. The results are also applicable to controlling technological performances in various contexts, such as arrays of mechanical resonators, catalytic surface reactions, and nonlinear optics.
	\end{abstract}
	
	\maketitle
	
	
	\textbf{Periodically forced oscillators can adjust their frequency to an integer fraction of the driving force over a wide range of parameters that are related to the applied forcing. This frequency-locking behavior generally displays a `V'-like resonant domain, a.k.a Arnol'd tongue, and is well understood in homogeneous media. Little is known, however, on resonance tongues in spatially heterogeneous media. Here we use a generic complex Ginzburg--Landau equation that is valid near the oscillation onset and show how spatial heterogeneity affects the resonance domain to form also `U' and `W' shapes. In addition, we uncover how these resonance properties change once the continuous limit is replaced by a chain of oscillators. The results are potentially applicable to systems with intrinsic monotonic spatial dependence or to systems in which spatial gradients or oscillating elements can be tailored to reach the desired performance. {Our results also apply to unforced coupled oscillatory media and shed light on the phenomenon of amplitude death and birth.}}
	
	\section{Introduction}
	
	Oscillatory media can be entrained to an external periodic driving force to exhibit resonant oscillations in which, the actual system's frequency is adjusted to an integer fraction of the driving frequency. In the vicinity of the oscillations onset, frequency-locking traditionally results in a `V' shape domain in the parameter plane spanned by the detuning and forcing amplitude~\cite{arnold2012geometrical}, also known as Arnol'd tongue. In spatially extended homogeneous media the leading Arnol'd tongues, e.g, 1:1 and 2:1 resonances, are the same as for a single oscillator. Yet, in some cases, spatial instabilities may break the spatial symmetry by forming patterns that change the Arnol'd tongue structure. e.g., extending frequency locking outside the Arnol'd tongue~\cite{yochelis2002development,castiloyochelis} or destroy entrainment inside the resonance region~\cite{yochelis2004frequency}.
	
	{Experimental studies of resonant oscillations in chemical, granular, optical and fluid systems, generally focused on spatially homogeneous conditions with driving forcing of additive and/or parametric nature~\cite{petrov1997resonant,lin2000resonant,melo1995hexagons,melo1994transition,arbell2000temporally,shats2012parametrically,coullet1994excitable,tlidi1998kinetics,izus2000bloch,imbihl1995oscillatory}. However, in some systems spatial inhomogeneity is an inherent feature of the system, such as the cochlea in the auditory system~\cite{gold1948hearing,dallos1996overview,hudspeth2010,AshmoreAvanBrownellEtAl2010,Hudspeth2014}, while in other systems inhomogeneities are exploited to tune the performance of a particular engineered output, such as mechanical resonators (NEMS and MEMS)~\cite{lifshitz2010nonlinear,jia2013parametrically,abrams2014nonlinear}, catalytic surface reactions~\cite{imbihl1995oscillatory}, coupled cavity soliton-based Kerr frequency combs~\cite{longhi1996stable,jang2019observation}, and Faraday-type waves~\cite{moisy2012cross,moriarty2011faraday,urra2019localized}. Yet, periodic forcing of inhomogeneous systems has been insufficiently analyzed~\cite{meron2015nonlinear}, hitherto.}
	
	In this study, we examine the effect of spatial heterogeneity in the natural frequency on the onset of resonant oscillations (hereafter `resonance onset' or `threshold') using the forced complex Ginzburg-Landau (FCGL) equation. {We focus on parametric forcing for which resonance tongues can be clearly defined as domains of resonant oscillations outside of which the oscillation amplitude is strictly zero. This is unlike additive forcing where small-amplitude resonant oscillations exist even far from resonance conditions. We refer the reader to Refs.~\cite{ourepl,ourPRE,ourPhysD} for more details about parametric vs. additive forcing.}
	We begin in section~\ref{sec:continuum} by demonstrating how selected spatial-heterogeneity forms lead in to resonance tongues of `U' and `W' shapes, rather than the commonly known `V' shape resonance tongue. Then in section~\ref{sec:chain}, we shift the focus to a discrete chain of oscillators and show how a decrease in the number of oscillators affects the resonance onset and frequency-locking boundaries. Finally, in section~\ref{sec:disc}, we summarize the results, {address the effects of other boundary conditions and spatial heterogeneities, and also relate the insights to previous works on unforced systems.}
	
	\section{Entrainment of parametrically driven oscillations with nonuniform frequencies}\label{sec:continuum}
	
\begin{figure}[tp]
		(a)\includegraphics[width=0.9\linewidth]{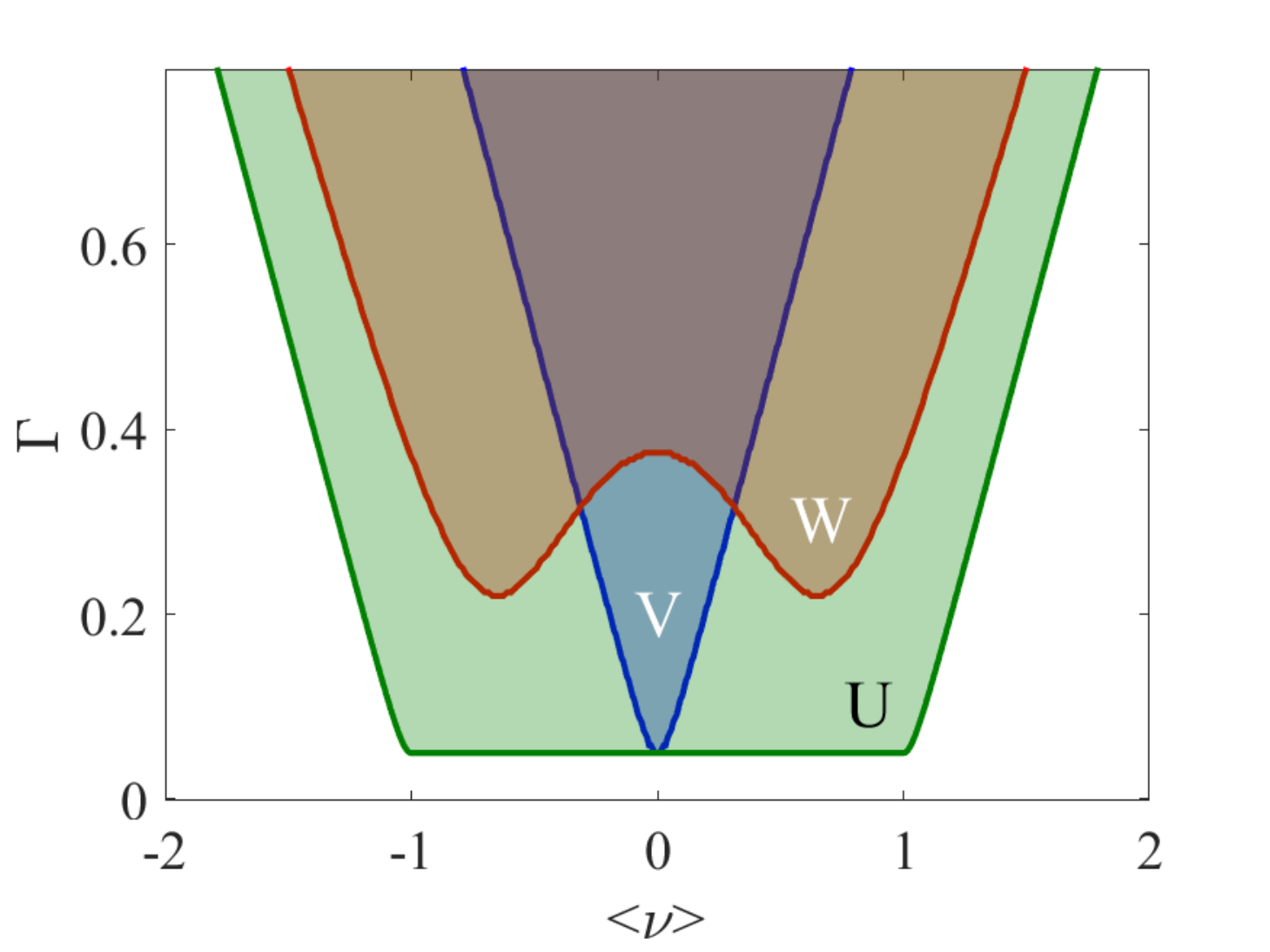}
		(b)\includegraphics[width=0.9\linewidth]{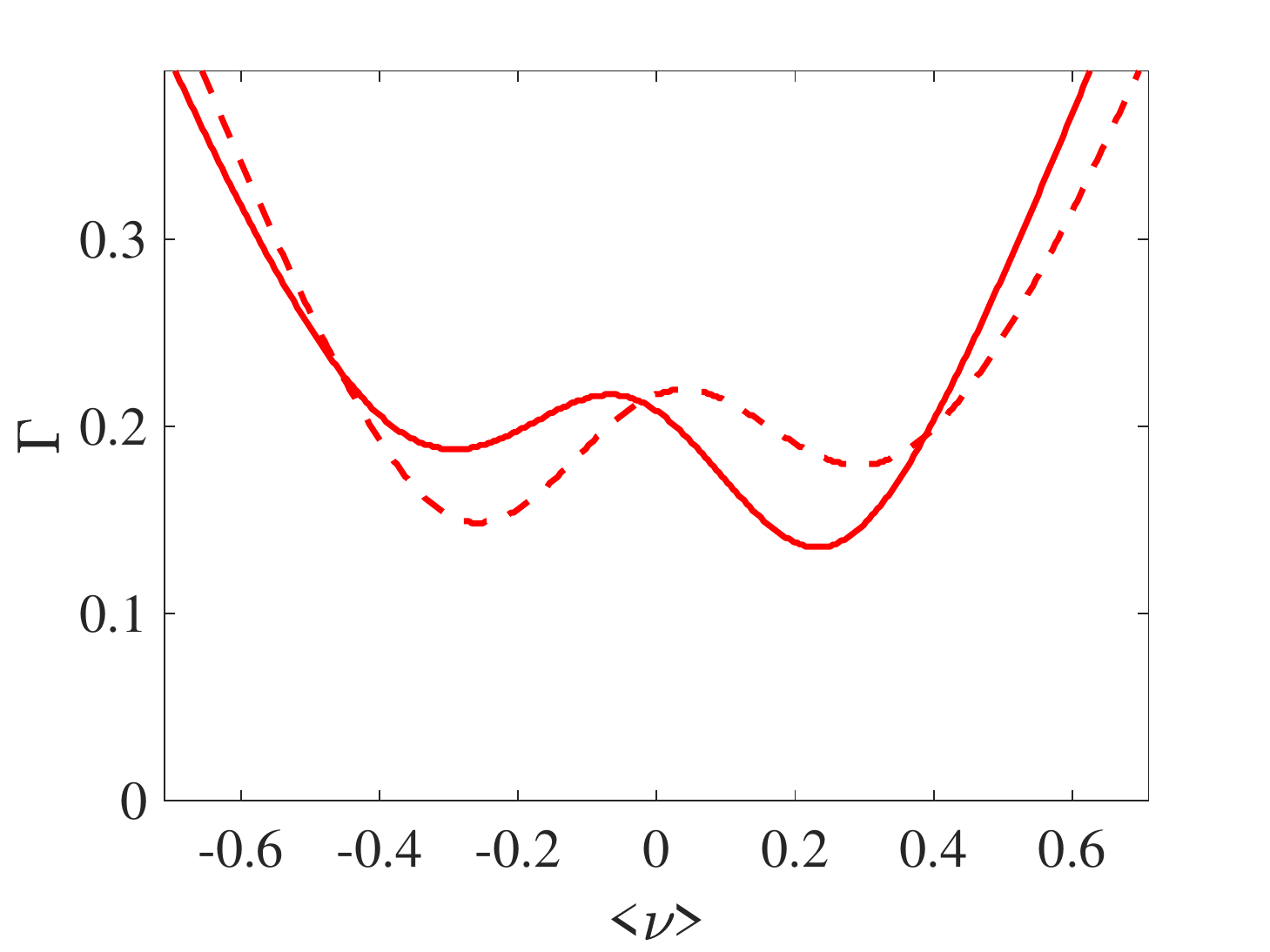}
		\caption{(a) Resonance tongues in a parameter plane of average detuning ($\avr{\nu}$) and forcing amplitude ($\Gamma$), obtained for~\eqref{eq:modelFCGL}: `V' form for homogeneous distribution of frequencies ($\eta=0$), `U' and `W' forms for a linear spatial dependence $\nu=\nu_L$ (see \eqref{eq:nu_L}), with $D=10^{-6}$ and $D=0.02$, respectively. The boundaries ($\Gamma_c$) for the homogeneous case (the `V' form) are given by~\eqref{eq:res_uni} while for the heterogeneous case, they are computed numerically using the eigenvalue problem~\eqref{eq:lns}. (b) Asymmetrical `W' shape resonance tongue for the convex (see~\eqref{eq:quad}) and concave forms (see~\eqref{eq:loghetro}) of $f(x)$, depicted by solid and dashed lines, respectively. Other parameters: $\mu=-0.05$.}\label{fig:newtongue}
	\end{figure}

	Frequency locking in a spatially extended oscillatory medium driven by parametric forcing is well described by the FCGL amplitude equation~\cite{CoulletEmilsson1992,pickles_book,meron2015nonlinear}, where near  1:1 and 2:1 resonance it reads as~\cite{lifshitz,ourepl}:
	\begin{equation}\label{eq:FCGL}
	\frac{\partial A}{\partial t}=\left(\mu+i\nu \right)A-(1+i\beta)|A|^2A+\Gamma \bar{A}+(1+i\alpha)\nabla^2A,
	\end{equation}
	where $\bar{A}$ is the complex conjugate of $A$, the parameters $\mu,\nu,\beta,\alpha \in \mathbb{R}$ describe, respectively, the distance from the Hopf bifurcation, detuning from the unforced frequency, nonlinear frequency correction and dispersion, and $\Gamma\in \mathbb{C}$ is related to the amplitude of the periodic parametric forcing. Near the Hopf bifurcation, the domain of any frequency-locked oscillations typically has a `V' shape in the forcing amplitude vs. detuning parameter plane, as shown equally for 1:1 or 2:1 resonances (according to~\eqref{eq:FCGL}) in Fig.~\ref{fig:newtongue}.
	
	In what follows, we focus on how spatial heterogeneity reshapes the resonance region for damped oscillations ($\mu<0$); specifically we use throughout all the computations $\mu=-0.05$. To do so, we consider a simplified version of~\eqref{eq:FCGL} in one spatial dimension on an interval of unit length ($x\in[-\sfrac{1}{2},\sfrac{1}{2}]$), impose monotonic inhomogeneity $\nu(x)=\eta f(x)+\nu_0$ and set $\beta=\alpha=0$ together with $0<\Gamma \in \mathbb{R}$:
	\begin{equation}\label{eq:modelFCGL}
	\frac{\partial A}{\partial t}=\left(\mu+i\nu(x) \right)A-|A|^2A+\Gamma\bar{A}+D\frac{\partial^2 A}{\partial x^2},
	\end{equation}
	{where $\nu_0$ is the detuning of the homogeneous system. The spatial inhomogeneity in $\nu$ shifts the minima of the resonance tongue. To keep consistency with the homogeneous system, we remove this shift by introducing an averaged detuning
		\begin{equation}
		\avr{\nu}=\nu_0+\eta<f(x)>=\nu_0+\eta\int_{-\frac{1}{2}}^{\frac{1}{2}}f(x)dx,
		\end{equation}
		so that a typical `V' shape always has a minimum at $\avr{\nu}=0$, as shown in Fig.~\ref{fig:newtongue}(a).} The choice of $\beta=0$ is made to simplify the discussion, as the bifurcation to resonant oscillations in this setting is supercritical~\cite{BurkeYochelisKnobloch2008}. Note that we introduced in~\eqref{eq:modelFCGL}, a diffusion scale $D$, which is scaled out in~\eqref{eq:FCGL}, to elaborate on the physical impact of spatial coupling in a system of fixed length.
	
	For $\eta=0$, the onset of frequency locking can be deduced from the dispersion relation~\cite{yochelis2002development} (linearization about the trivial solution to~\eqref{eq:modelFCGL}):
	\begin{equation}
	\nonumber \sigma(k)=\mu-Dk^2+\sqrt{\Gamma^2-\nu_0^2}.
	\end{equation}
	Since the Hopf bifurcation occurs at $\sigma(k=0)=0$, the critical amplitude of the forcing that is needed to attain entrainment, is given by
	\begin{equation}\label{eq:res_uni}
	\Gamma_c=\sqrt{\mu^2+\nu_0^2},
	\end{equation}
	which forms the the typical `V' form, as shown in Fig.~\ref{fig:newtongue}(a).
	
	However, once spatial monotonic heterogeneity is introduced ($\eta\neq0$), the resonance tongue can demonstrate, under certain choices of $\eta$, also other shapes: an ``inverse camel'' shape (or `W') and a ``one-half barrel'' shape (or `U'), as shown in Fig.~\ref{fig:newtongue}(a). Particularly, the distinct `W' form implies that systems with monotonic spatial distributions of frequencies ($\eta\ne 0$) favor locking away from the natural frequency of the homogeneous system ($\eta=0$), i.e., ($\avr{\nu}\ne 0$), as periodic forcing is applied and moreover, that the asymmetry of the shape can be tuned by changing the function $f(x)$ from a convex to a concave form. To demonstrate the symmetry breaking in $\avr{\nu}$, we use a convex choice
	\begin{equation}\label{eq:quad}
	f(x)=(x+1/2)^2,
	\end{equation}
	where the well at $\avr{\nu}>0$ is deeper, and a concave choice
	\begin{equation}\label{eq:loghetro}
	f(x)=\log\left(x+1\right),
	\end{equation}	
	where the well at $\avr{\nu}<0$ is deeper. Figure~\ref{fig:newtongue}(b) depicts the resonance tongues for both cases.
	
	In what follows, we consider first a continuous limit and address the mechanisms that create the symmetric resonance tongue employing linear form of $f(x)$ (see Fig.~\ref{fig:newtongue}(a)),
	\begin{equation}\label{eq:nu_L}
	\nu_L(x)=2\eta x+\nu_0,
	\end{equation}
	for which $\avr{\nu}=\nu_0$. We then consider a discrete chain of oscillators and discuss similarities and dissimilarities with respect to the continuous limit.
	
	\subsection{Effect of spatial heterogeneity}
	We consider the linear spatial distribution of frequencies given by Eq.~\ref{eq:nu_L}, which is easier to analyze. To obtain the onset of resonant oscillations, we study the linear stability of the trivial solution, $A=0$, of~\eqref{eq:modelFCGL} to infinitesimal perturbations of the form:
	\begin{equation}\label{eq:lin_anal}
	A\simeq a(x)e^{\lambda t},
	\end{equation}
	where $\lambda$ is the eigenvalue that indicates the growth rate of the eigenfunction $a(x)$. Inserting~\eqref{eq:lin_anal} into~\eqref{eq:modelFCGL} and keeping leading order contributions, we obtain the characteristic equation
	\begin{equation}\label{eq:lns}
	\lambda a=\left(\mu+i\nu_L\right)a+\Gamma \bar{a}+Da'',
	\end{equation}
	where primes denote spatial derivatives. Although~\eqref{eq:lns} resembles a modified version of the complex Airy equation~\cite{vallee2004airy}, the addition of the parametric forcing precludes an analytic treatment. We therefore solve~\eqref{eq:lns} numerically using Neumann boundary conditions.
	
\begin{figure}[tp]
		\includegraphics[width=0.9\linewidth]{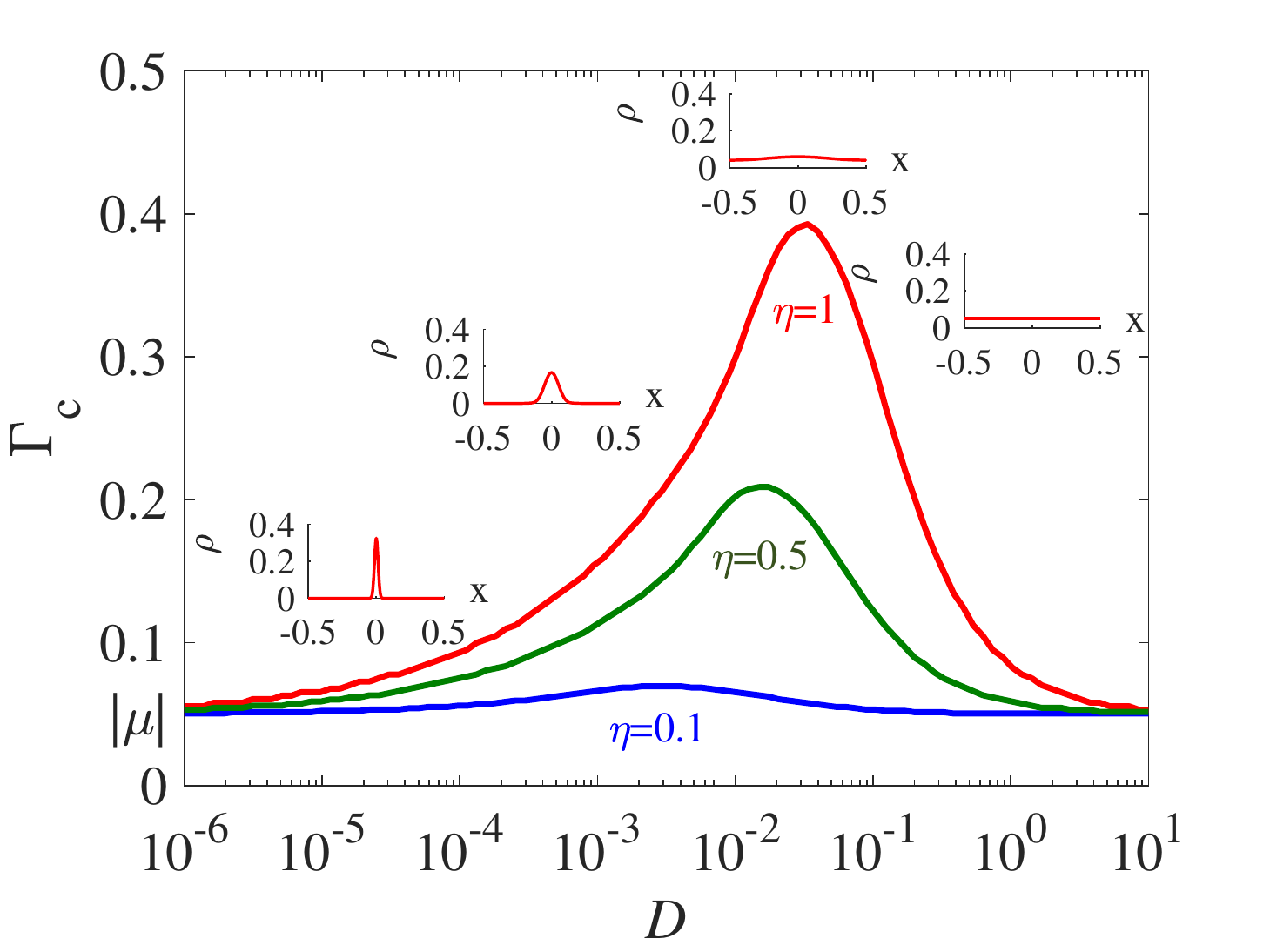}
		\caption{Resonance onset ($\Gamma_c$) as a function of spatial coupling strength ($D$) at several $\eta$ values and $\nu_0=0$. The insets show the amplitudes of the eigenfunctions $a(x)=\rho \exp(i\phi)$ that correspond to the critical eigenvalue $\lambda_c=0$, at different values of $D=10^{-6},10^{-4},0.03,10$ along the curve $\eta=1$. The results were obtained numerically by solving~\eqref{eq:lns}.}\label{fig:theEffectOfEta}
	\end{figure}

	We start by calculating the instability threshold, $\Gamma_c$, that leads to resonant oscillations, varying $D$ for a few representative values of $\eta$, assuming $\nu_0=0$, as shown in Figure~\ref{fig:theEffectOfEta}. To this end, we set the eigenvalue to zero, $\lambda=\lambda_c=0$, and identify the value of $\Gamma$ for which a solution of~\eqref{eq:lns} satisfying the boundary conditions, exists. That solution represents the eigenfunction associated with $\lambda_c=0$.
	The insets in Figure~\ref{fig:theEffectOfEta}, show the respective eigenfunctions along the curve obtained for $\eta=1$. Notably, near the onset, the eigenfunctions provide good approximations to the actual solutions of~\eqref{eq:modelFCGL}. Unlike the case of a homogeneous system, the resonant-oscillations threshold, strongly depends on the spatial coupling. Specifically, the threshold shows a hump-type behavior in $\Gamma_c$: First the onset increases with $D$ from $\Gamma_c = |\mu|$ and then decreases as $D\to\infty$ back to $\Gamma_c= |\mu|$. 
	
	For $D\to0$, the spatial coupling is negligible and the system behaves like a continuum of independent oscillators, parametrized by $x$. Each oscillator has its own natural frequency detuned by $\nu_L(x)$, and oscillation threshold $\Gamma_c(x)$, obtained by solving~\eqref{eq:lns} with $\lambda=D=0$. The oscillation threshold for the whole system and the first oscillator, $x_0$, to pass that threshold, can be obtained by minimizing $\Gamma_c(x)$ with respect to $x$. This gives $\Gamma_c=|\mu|$ and $x_0=-\nu_0/(2\eta)$. Thus, near the onset only a small fraction of the domain is entrained, and for $\nu_0=0$ that fraction is centered around $x_0=0$, as indicated by the spatially localized amplitude of the eigenfunction in the left-most inset of Fig.~\ref{fig:theEffectOfEta}. By contrast, for $D\to\infty$ the spatial coupling is strong and the whole domain is synchronized at the same amplitude, as the right-most inset in Fig.~\ref{fig:theEffectOfEta} shows. Mathematically, the strong-coupling limit implies
	\begin{equation}\label{stronglimit}
	\lim_{D\to\infty} a'' = 0
	\end{equation}
	so that the solutions to~\eqref{eq:lns} (satisfying the Neumann boundary conditions) are simply $a(x)=\text{constant}$. The system then behaves as a uniform oscillating medium with a natural frequency detuned by $\avr{\nu}=\nu_0=0$ and oscillation threshold $\Gamma_c=|\mu|$ as Eq.~\ref{eq:res_uni} implies. Consequently, in both limits $D\to 0$ and $D\to\infty$ for $\nu_0=0$, the onset is independent of $\eta$ and approaches asymptotically the value  $\Gamma_c=|\mu|$.
	
	In between the two limiting cases for $D$, the oscillation threshold increases with $\eta$ in a nonlinear (albeit monotonic) fashion. This behavior, in fact, is related to the phenomena of ``oscillator death'' that is observed in unforced oscillations~\cite{kedma1985,aronson1990amplitude,koseska2013oscillation}. In this situation, coupled oscillators may have a destructive dynamics that lead to the overall damping, although they are all above the onset of the Hopf bifurcation. Such damping can result, for example, from delayed coupling~\cite{reddy1998time} or nonidentical properties, such as different natural frequencies~\cite{aronson1990amplitude}. The latter is analogous to the space-dependent detuning considered here. An increase in the coupling strength $D$ widens the spatial domain of resonance, due to stronger interactions between nearby oscillations, as Fig.~\ref{fig:theEffectOfEta} shows. Since the detuning is not identical, these interactions are destructive and thus, require higher forcing strength for entrainment. Hence, the threshold, $\Gamma_c$, increases with $D$. Similarly, an increase in heterogeneity (here $\eta$) also results in increased threshold as it is analogous to increasing the frequency difference between adjacent points of oscillations. For $D\to\infty$ the destructive effects become again negligible due to synchronization both in phase and in amplitude, which agrees with the resonance onset decrease toward $\Gamma_c\to |\mu|$.
	
	In what follows, we use the above properties to gain insights into the distinct forms of the resonance tongue, i.e., for $\nu_0 \neq 0$.
	
	\subsection{Effect of detuning on Arnold's tongues}
	
\begin{figure}[tp]
		\includegraphics[width=0.9\linewidth]{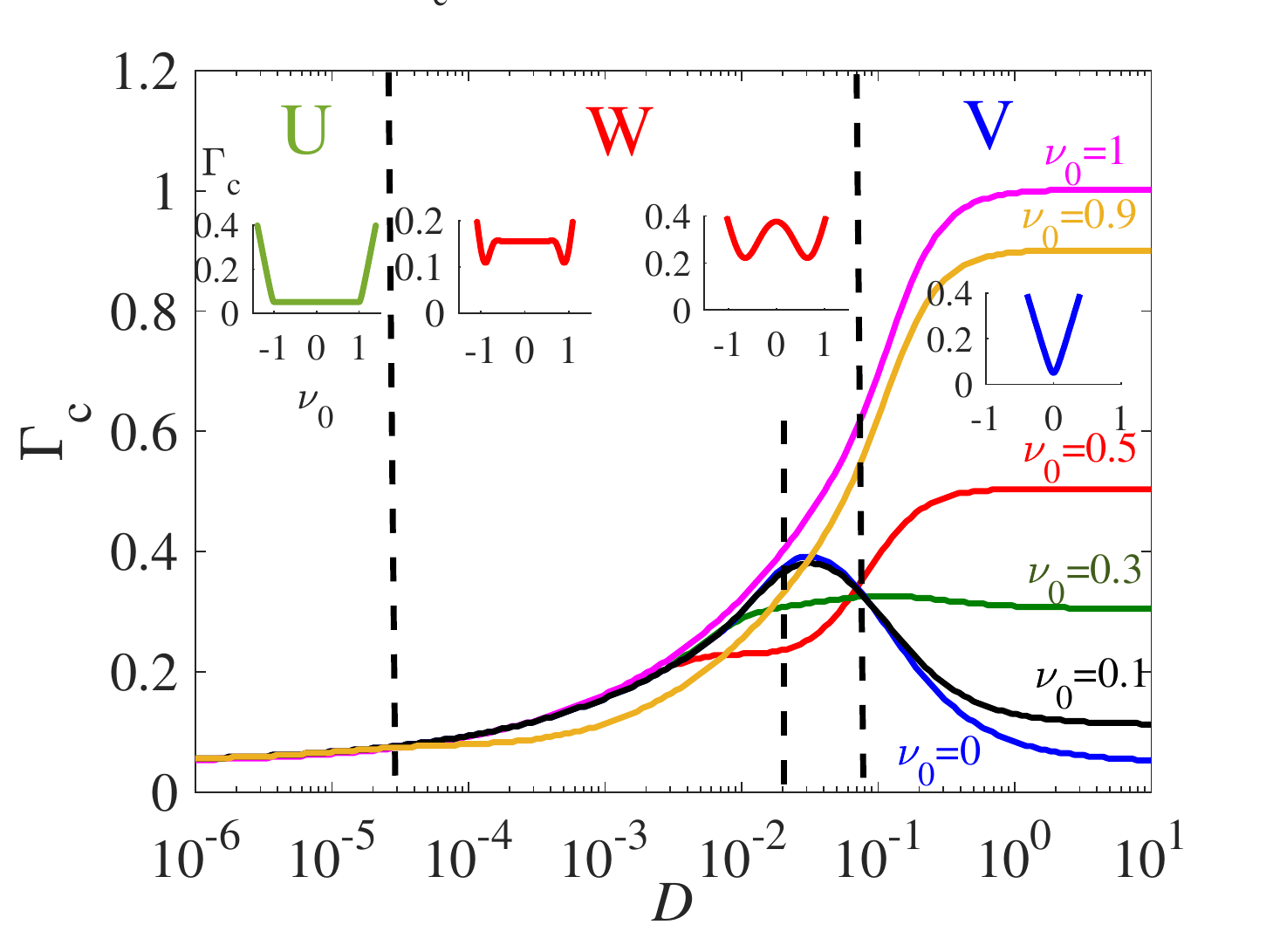}
		\caption{Resonance onset ($\Gamma_c$) as a function of spatial coupling strength ($D$) at several $\nu_0$ values while keeping $\eta=1$; the results were obtained numerically by solving~\eqref{eq:lns}. Top insets represent typical resonance tongue forms in between the vertical dashed lines, respectively.
		}\label{fig:theEffectOfnu0}
	\end{figure}

	We study the impact of nonzero average detuning, that is $\avr{\nu_L}=\nu_0$, considering first the two asymptotic limits $D \to 0$ and $D \to \infty$. We then examine intermediate $D$ values, which lead to the `W' shape of the resonance tongue, as shown in Fig.~\ref{fig:theEffectOfnu0}.
	
\begin{figure*}[tp]
		(a)\includegraphics[width=0.43\linewidth]{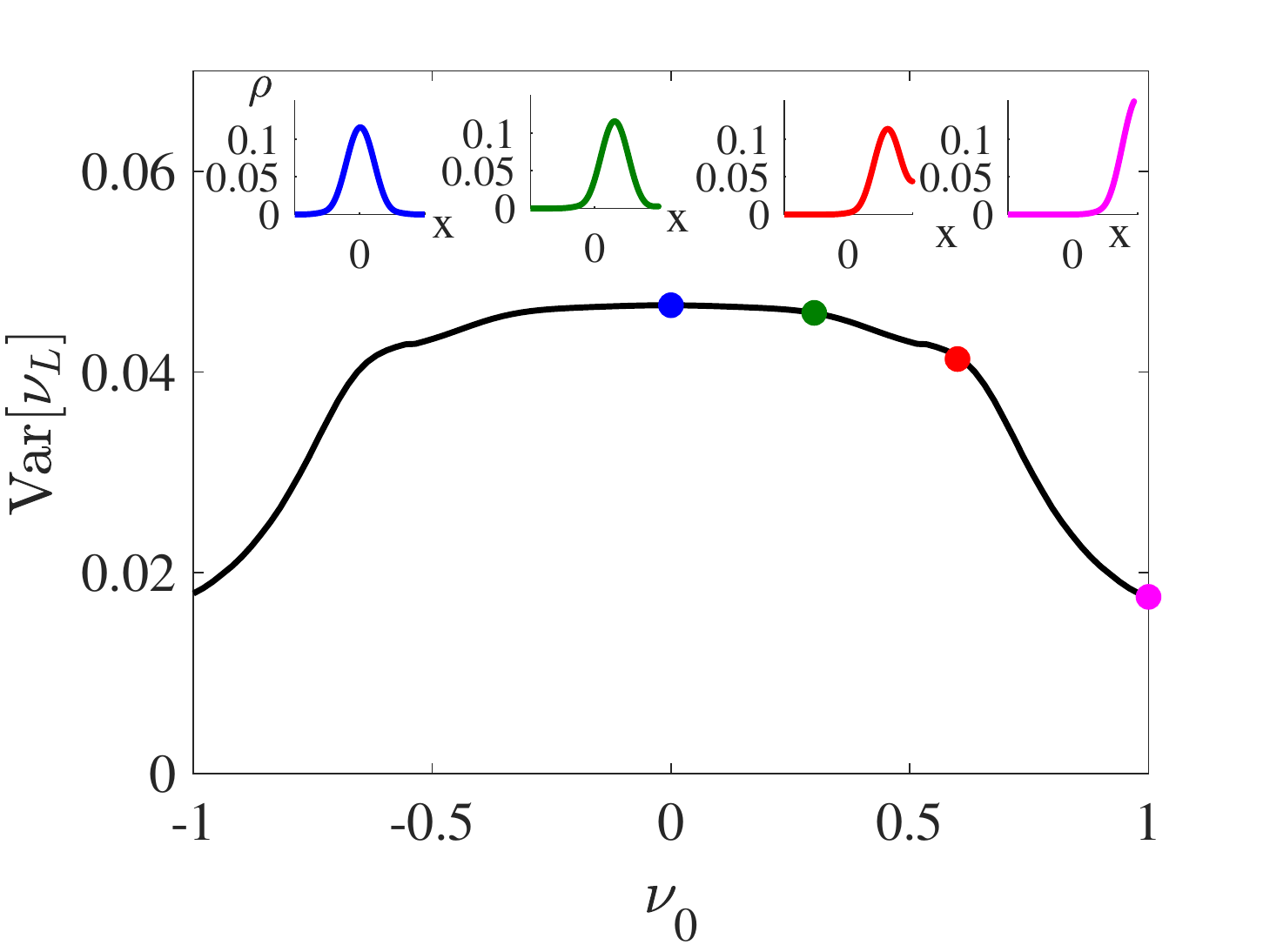}
		(b)\includegraphics[width=0.43\linewidth]{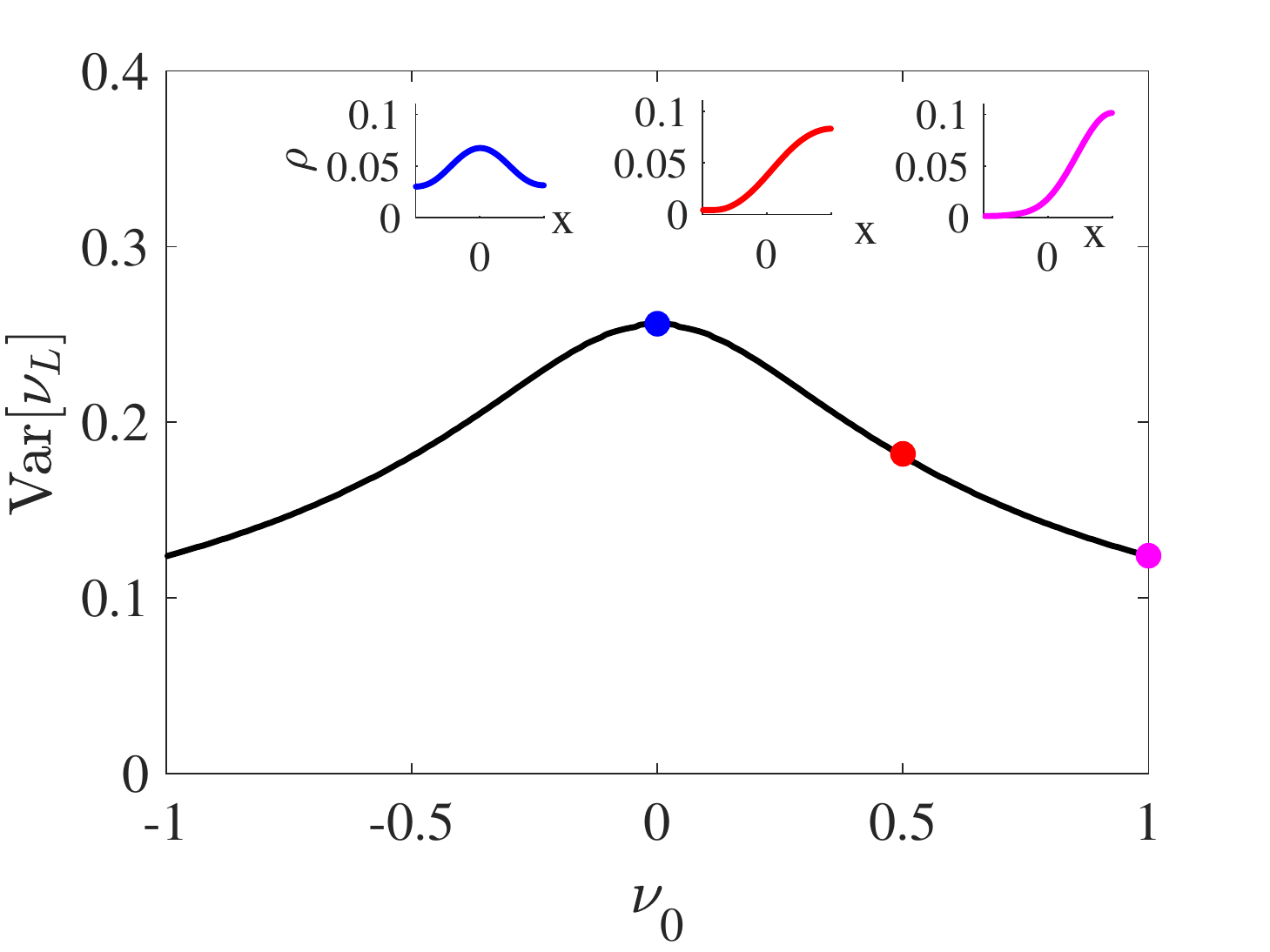}
		\caption{(a) Variance of linear detuning $\nu_L$ as a function of $\nu_0$ for $D=10^{-3}$ that is within the 'W' shape regime, obtained using~\eqref{eq:varnu}. (b) Variance for $D=2\cdot 10^{-2}$. The insets show the eigenfunction profiles in terms of the amplitude $\rho(x)$, at marked locations, respectively: (a) $\nu_0=0,0.3,0.6,1$ and (b) $\nu_0=0,0.5,1$.}\label{fig:varnu}
	\end{figure*}

	In the vanishing-coupling limit, $D\to 0$, the eigenfunction, $a(x)$, is spatially localized, and for $|\nu|<|\eta|$, the location of its hump is shifted to $x_0=-\nu_0/(2\eta)$ (a condition for which $\nu_L=0$). Since $\Gamma_c=|\mu|$ is independent of $\nu_0$ and $\eta$, as long as $|\nu_0|<|\eta|$, it is independent of the hump location $x_0$, implying equal oscillations amplitudes $\rho=|a|$ for different $\nu_0$ values, as the insets in Fig.~\ref{fig:varnu}(a) show. Otherwise the hump will always be located at the boundary, i.e., at $x_0=\pm 1/2$.
	When the hump is pinned to the boundary the threshold increases with $\nu_0$ according to~\eqref{eq:res_uni}, which is the typical `V' shape tongue. This behavior yields the 'U' shape tongue, as is also shown in Fig.~\ref{fig:newtongue}(a):
	\begin{equation}\label{eq:gam_nu0}
	\Gamma^{{D \to 0}}_c  = \left\{ {\begin{array}{cl}
		{|\mu| ,} & \text{and}\quad |\nu_0|<|\eta| \\ \\
		{\sqrt{\mu^2+(|\eta|-|\nu_0|)^2} ,} &  \text{and}\quad|\nu_0|\ge |\eta|.
		\end{array}} \right.
	\end{equation}
	On the contrary, in the strong coupling limit ($D\to\infty$), the domain oscillates uniformly and thus the resonance onset is the same as for a single oscillator with detuning $\nu_0$ and the resonance domain assumes the typical `V' shape resonance.
	
	In the intermediate coupling regime, about $10^{-5}<D<10^{-1}$, the tongue displays a non-monotonic `W' form that is different from the 'V' or 'U' shapes, see Fig.~\ref{fig:theEffectOfnu0}. In this regime, the minimal $\Gamma_c$ value for resonance is no longer at $\nu_0=0$ (as for the `V' form), but at finite $\nu_0$ which depends on $D$. Moreover, the `W' shape range of $D$ is divided into two sub-ranges: For relatively small D values the tongue displays a plateau in the vicinity of $\nu_0=0$, while for relatively large D values, $\Gamma_c$ has an extremum point at $\nu_0\ne 0$.
	
	To unfold the distinct features inside the `W' shape region, we find it useful to define a measure for the effective heterogeneity level, which is a variance of the detuning, weighted by the amplitude of the eigenfunction associated with the zero eigenvalue::
	\begin{equation}\label{eq:varnu}
	\Var[\nu_L]=\int_{-\frac{1}{2}}^{\frac{1}{2}}\nu^2_L(x)\hat \rho(x)dx-\left[\int_{-\frac{1}{2}}^{\frac{1}{2}}\nu_L(x)\hat \rho(x)dx\right]^2,
	\end{equation}
	where $\hat \rho(x)=\rho(x)/\int\rho(x)dx$ is the normalized amplitude of the eigenfunction.
	
	{The `W' shape resonance in the low (left) sub-range of $D$ includes a plateau, which has the same origin as the plateau in the `U' shape: Increase of $|\nu_0|$ shifts the location of the eigenfunction's hump but as long as the tail vanishes at the boundary, the changes in the threshold $\Gamma_c$ and in the variance are negligible, see~Fig.~\ref{fig:theEffectOfnu0} and~\ref{fig:varnu}(a), respectively. When the eigenfunction is shifted further to the boundary ($\nu_0\to \eta$), the eigenfunction becomes narrower which is indicated also by lower variance value; see inset associated with $\nu_0=0.6$ in Fig.~\ref{fig:varnu}(a). Decrease in the variance means that there is less competition between oscillations, which in turn makes the entrainment easier to reach, i.e., to obtain a lower resonance threshold and thus forming a minima in the `W' shape, see the curve for $\nu_0=0.9$ in Fig.~\ref{fig:theEffectOfnu0}. Finally, as in the `U' shape resonance mechanism, when $\nu_0$ is increased above a certain value, it becomes harder to overcome the difference between the natural oscillation and the the driving force, thus, entrainment requires increase in $\Gamma_c$.
		
	The right sub-region (larger $D$ values) of `W' shape is in general similar to left sub-region but with one distinction: At $\nu_0=0$, the amplitudes of the eigenfunction do not vanish ($\rho(x) > 0$) at $x= \pm 1/2$, see the right inset in Fig.~\ref{fig:varnu}(b). In this case, the hump is no longer invariant (keeps the same form) since any shift in the location (by changing $\nu_0$) results immediately in a narrower eigenfunction and so respectively in both lower variance and onset.	As such, the right sub-region of the `W' shape is a finite size effect. As in the previous case, the `V' shape tongue is retained for large enough $\nu_0$ values.}
	
	Next, we turn to examine the deviation of the results obtained above when the continuous medium is replaced by a discrete chain of oscillators.
	
	\section{From a continuous limit to a chain of coupled oscillators}\label{sec:chain}
	
	To analyze a chain with $N$ oscillators, we rewrite~\eqref{eq:modelFCGL} in a discrete form for each individual oscillator:
	\begin{equation}\label{eq:discrete}
	\begin{split}
	\frac{\partial A_j}{\partial t}=&\left(\mu+i\nu_L^j \right)A_j-|A_j|^2A_j+\Gamma \bar{A}_j+\\
	&\frac{D}{(N-1)^2}(A_{j+1}-2A_j+A_{j-1}),
	\end{split}
	\end{equation}
	where the monotonic variation in frequency along the chain is given by
	\begin{equation}\label{eq:disc_nu}
	\nu^j_L=\eta\left(\frac{2(j-1)}{N-1}-1\right)+\nu_0,
	\end{equation}
	with $j=1,2,\dots,N$. We note that normalization factor $(N-1)^2$ is analogues to the spatial grid refining that is used in finite difference method for numerical integrations of parabolic partial differential equations, i.e., for a chain of unity length, $N\to \infty$ is equivalent to $\Delta x\to 0$. The Neumann boundary conditions are retained in a standard fashion, by setting $A_0=A_1$ and $A_{N}=A_{N+1}$.
	
	Linear stability analysis of~\eqref{eq:discrete} around the zero state yields a Jacobian matrix that can be written as
	\begin{equation}\label{eq:JAC}
	J=\left(\begin{array}{cc}
	M_+ & -L \\
	L & M_-
	\end{array}\right)
	\end{equation}
	where $M,\,L$ are sparse $N\times N$ matrices that include also the Neumann boundary conditions:
	\begin{widetext}
		\[
		M_\pm=\left(\begin{array}{cccccc}
		\mu\pm\Gamma-D & D &  \\
		D & \mu\pm\Gamma-2D & \ddots  \\
		& \ddots & \ddots & \ddots \\
		& & \ddots & \mu\pm\Gamma-2D & D\\
		& & & D & \mu\pm\Gamma-D
		\end{array}\right),\quad
		L=\left(\begin{array}{cccccc}
		\nu^1_L & &  \\
		& \ddots &   \\
		&  & \ddots &  \\
		& &  & \ddots & \\
		& & &  & \nu^{N}_L
		\end{array}\right).
		\]
	\end{widetext}
	
\begin{figure}[tp]
		\includegraphics[width=0.9\linewidth]{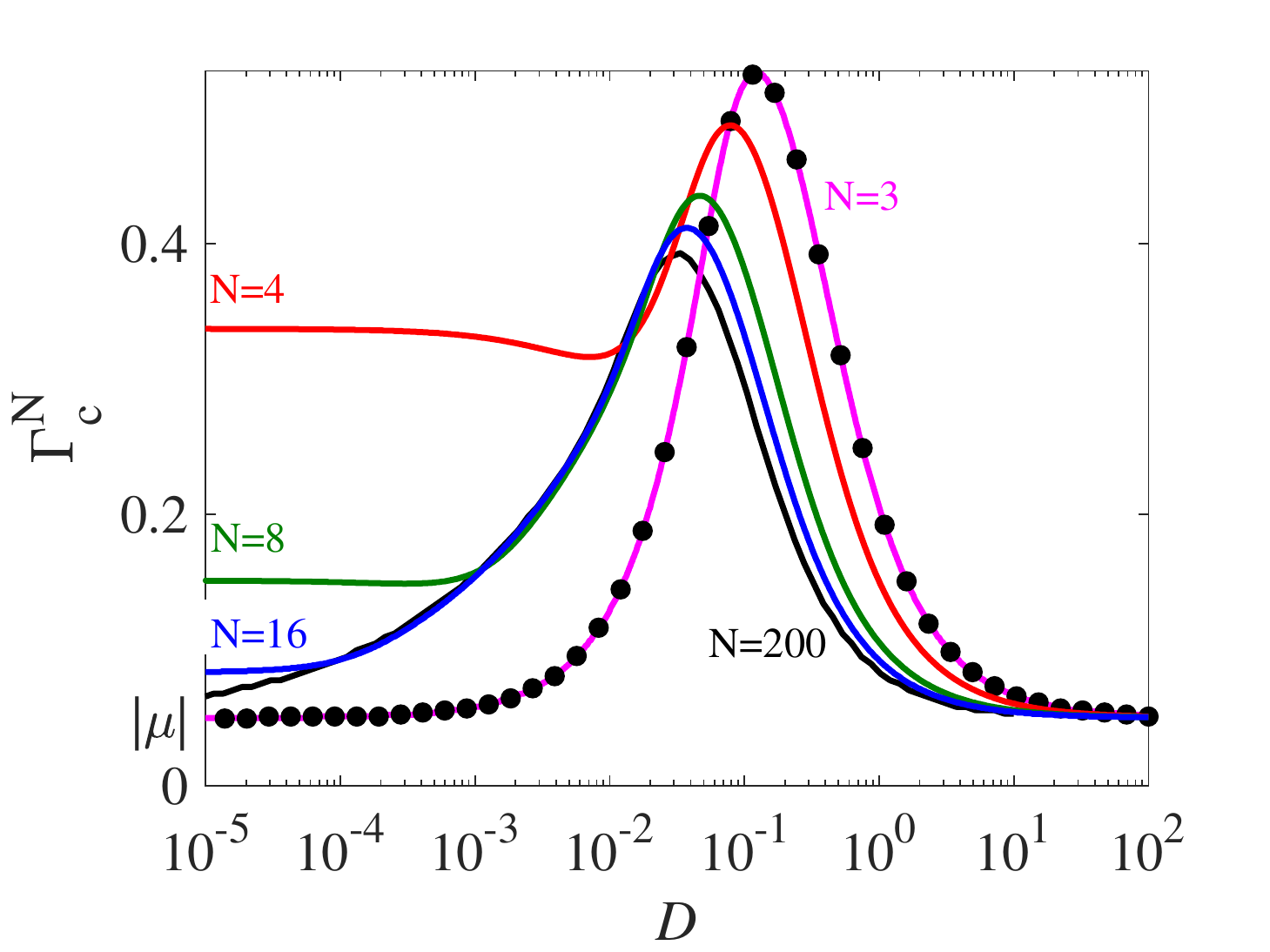}
		\caption{Resonance onset ($\Gamma^N_c$) as a function of spatial coupling strength ($D$) at several number of oscillators $N$ while keeping $\eta=1$ and $\nu_0=0$; the results were obtained numerically using the Jacobian matrix~\eqref{eq:JAC}. For $N=3$, we plot the onset using the derived form for $\lambda_-$, i.e.,~\eqref{3osceigenval}, while the numerical eigenvalue computations are marked by `$\bullet$'.}
		\label{fig:diffn}
	\end{figure}

	In Figure~\ref{fig:diffn}, we show the main qualitative difference that is obtained (via stability analysis) already for $\nu_0=0$, demonstrated for $\eta=1$ (see Fig.~\ref{fig:theEffectOfEta}). Apart from a quantitative deviation obtained with decreasing $N$, in comparison to the curve continuous limit (here approximated with $N=200$), there is a distinct qualitative difference, which appears as a plateau in the vanishing coupling limit $D\to0$, only for even $N$ values. In what follows, we first explain this qualitative difference for even $N$ values ($N>2$). We then analyze in greater detail the simplest cases of even and odd numbers of oscillators, that is, $N=2$ and $N=3$.
	
	At the limit of vanishing $D$ values, the oscillators are decoupled and for odd $N$ there always exists an oscillator that is entrained exactly at $\nu=0$, and thus, the resonance criteria~\eqref{eq:res_uni} holds. For even $N$ values the condition $\nu=0$ lies always in between oscillators, so that the onset is determined by the adjacent two oscillators that are closest to the resonance, i.e., $j=\lfloor\frac{N+1}{2}+\frac{\nu_0(N-1)}{2\eta} \rceil$, where $\lfloor$ and $\rceil$ represent the lower- and upper-bound integers, respectively. Consequently, the resonance onset for $D\to0$ (Fig. \ref{fig:diffn}), follows:
	\begin{equation}\label{oddevenOnset}
	\Gamma^N_c  = \left\{ {\begin{array}{cl}
		{|\mu| ,} & {\text{for odd }N}  \\ \\
		{\sqrt{\mu^2+\dfrac{\eta^2}{N-1}} ,} & {\text{for even }N.} \\
		\end{array}} \right.
	\end{equation}
	Notably, the continuous limit of $\Gamma_c$ at finite $\eta$~\eqref{eq:gam_nu0}, is retained for $N\to\infty$.
	
	\subsection{Two coupled oscillators}
	
\begin{figure}[tp]
		\includegraphics[width=0.9\linewidth]{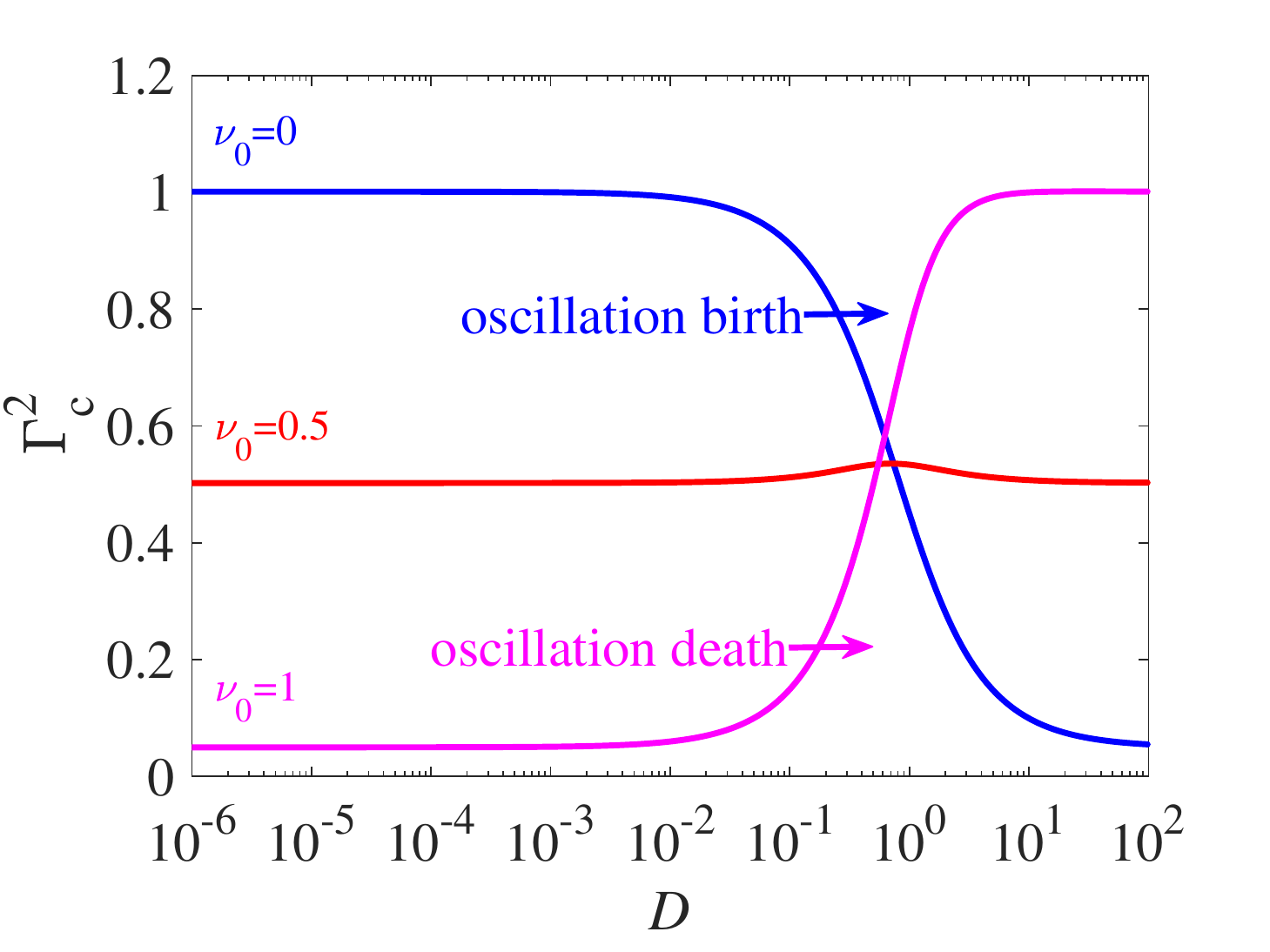}
		\caption{Resonance onset ($\Gamma_{c,2}$) for two coupled oscillators ($N=2$), according to~\eqref{eq:2osc}, as a function of spatial coupling strength ($D$) at several $\nu_0$ values, while keeping $\eta=1$. The term ``oscillation death'' (``oscillation birth'') refers to a change in $\Gamma_c$ , as $D$ is increased, that exceeds (falls short of) the forcing amplitude $\Gamma$, and, thus, leads to the destruction (generation) of oscillations.}
		\label{fig:osc2diffnu0}
	\end{figure}

	Coupling between two oscillators ($N=2$) is a special and distinct case. Yet, this case is amenable to analytic analysis using linear stability~\cite{pisarchik2003oscillation} that sheds light also on the continuous limit. The Jacobian matrix for two oscillators can be arranged in a simple form as:
	\[
	J_{2}=\left(\begin{array}{cccc}
	\mu+\Gamma-D & -\eta-\nu_0 & D & 0 \\
	\eta+\nu_0 & \mu-\Gamma-D & 0 & D\\
	D & 0 & \mu+\Gamma-D & \eta+\nu_0\\
	0 & D & -\eta-\nu_0 & \mu-\Gamma-D
	\end{array}\right),
	\]
	where the resulting eigenvalues are
	\[
	\begin{split}
	\lambda^\pm_{\pm}=\mu-D\pm \Big(&D^2-\eta^2+\Gamma^2-\nu_0^2\\
	&\pm 2\sqrt{D^2\Gamma^2-D^2\nu_0^2+\eta^2\nu_0^2}\,\Big)^{1/2}.
	\end{split}
	\]
	Among the four eigenvalues the critical one is $\lambda^+_+$, yielding the resonance onset:
	\begin{equation}\label{eq:threshold2Osc}
	\begin{split}
	\Gamma_{c,2}=\Big(&D^2+(\mu-D)^2+\eta^2+\nu_0^2\\
	&- 2\sqrt{D^2(\mu-D)^2+D^2\eta^2+\eta^2\nu_0^2}\,\Big)^{1/2}.
	\end{split}
	\end{equation}
	The analytic form admits two limiting behaviors with respect to $D$ and $\nu_0$
	\begin{equation}\label{eq:2osc}
	\Gamma_{c,2}  = \left\{ {\begin{array}{cl}
		{\sqrt{\mu^2+(|\eta|-|\nu_0|)^2},} & \text{for }D\to 0  \\ \\
		{\sqrt{\mu^2+\nu_0^2},} & \text{for }D\gg \mu,\eta,\nu_0 \\
		\end{array}}, \right.
	\end{equation}
	which agree with the uncoupled~\eqref{eq:gam_nu0} and the fully synchronized~\eqref{eq:res_uni} cases, respectively. The onset~\eqref{eq:2osc} indicates two asymptotic behaviors with a transition at $\nu_0=\eta/2$, in which for $\nu_0<\eta/2$ the onset for weak coupling is higher than for the strong coupling and vise versa, as shown in Fig.~\ref{fig:osc2diffnu0}. The biasymptotic behavior for $\nu_0>\eta/2$ is known as ``oscillation death''~\cite{pisarchik2003oscillation} that is increase in coupling strength leads to damping of both oscillators. Respectively, for $\nu_0<\eta/2$, we witness a reciprocal behavior to which we refer as ``oscillation birth'', namely, increase in coupling leads to revival of the oscillations at larger $D$ values.
	
	\subsection{Three coupled oscillators}
	
	The three oscillators case, is the maximal number of oscillators that is amenable to analytic treatment from which the hump-type form for $\Gamma_c$ as a function of $D$ can be obtained. The eigenvalue expressions are cumbersome, but it is possible to manipulate them numerically to find $\Gamma_c$ and graphically portrait the results, as shown in Fig.~\ref{fig:diffn}.
	
	Using the Jacobian~\eqref{eq:JAC}, we obtain the following eigenvalues:
	\begin{eqnarray}\label{3osceigenval}
	\lambda_{\pm} &=&\mu-\frac{4D\pm\Gamma}{3}+\frac{1}{3}\left(\frac{U_{\pm}}{S_{\pm}^{1/3}}+S_{\pm}^{1/3}\right), \\
	\lambda^q_{\pm} &=&\mu-\frac{4D\pm\Gamma}{3}- \\
	\nonumber &&\frac{1}{6}\left[\frac{U_{\pm}}{S_{\pm}^{1/3}}+S_{\pm}^{1/3}-(-1)^q\sqrt{3}i\left( \frac{U_{\pm}}{S_{\pm}^{1/3}}\pm S_{\pm}^{1/3}\right)\right],
	\end{eqnarray}
	where $q=0,1$, $U_{\pm}=7D^2\pm2D\Gamma+4\Gamma^2-3\eta^2$, and
	\small
	\[
	\begin{split}
	S_{\pm}=&\sqrt{3}\Big[-9D^6\pm18D^5 \Gamma+D^4 (23\eta^2+27\Gamma^2 )\\
	&-6D^3 (\mp 7\eta^2 \Gamma\pm6\Gamma^3 )+D^2 (-4\eta^4+42\eta^2 \Gamma^2-36\Gamma^4 )\\
	&+D(\pm4\eta^4 \Gamma\mp 4\eta^2 \Gamma^3 )+\eta^4 (\eta^2-\Gamma^2)\Big]^{1/2}\\
	&-10D^3\mp39D^2\Gamma+D(6\Gamma^2-9\eta^2)+\Gamma(\mp9\eta^2\pm8\Gamma^2 ).
	\end{split}
	\]
	\normalsize
	The resonance onset is determined by $\lambda^+_+$ and not only that it biasymptotes to $|\mu|$ as $D\to0$ and $D\to\infty$ but it captures the hump-like behavior with an excellent agreement to numerical computations, as shown in Fig.~\ref{fig:diffn}. This qualitative behavior holds for all odd number of oscillators and approaches quantitatively the continuous limit exactly as in the even $N$ case.
	
	\subsection{Resonance tongues for the discrete system}\label{sec:asym}
	
\begin{figure}
		(a)\includegraphics[width=0.9\linewidth]{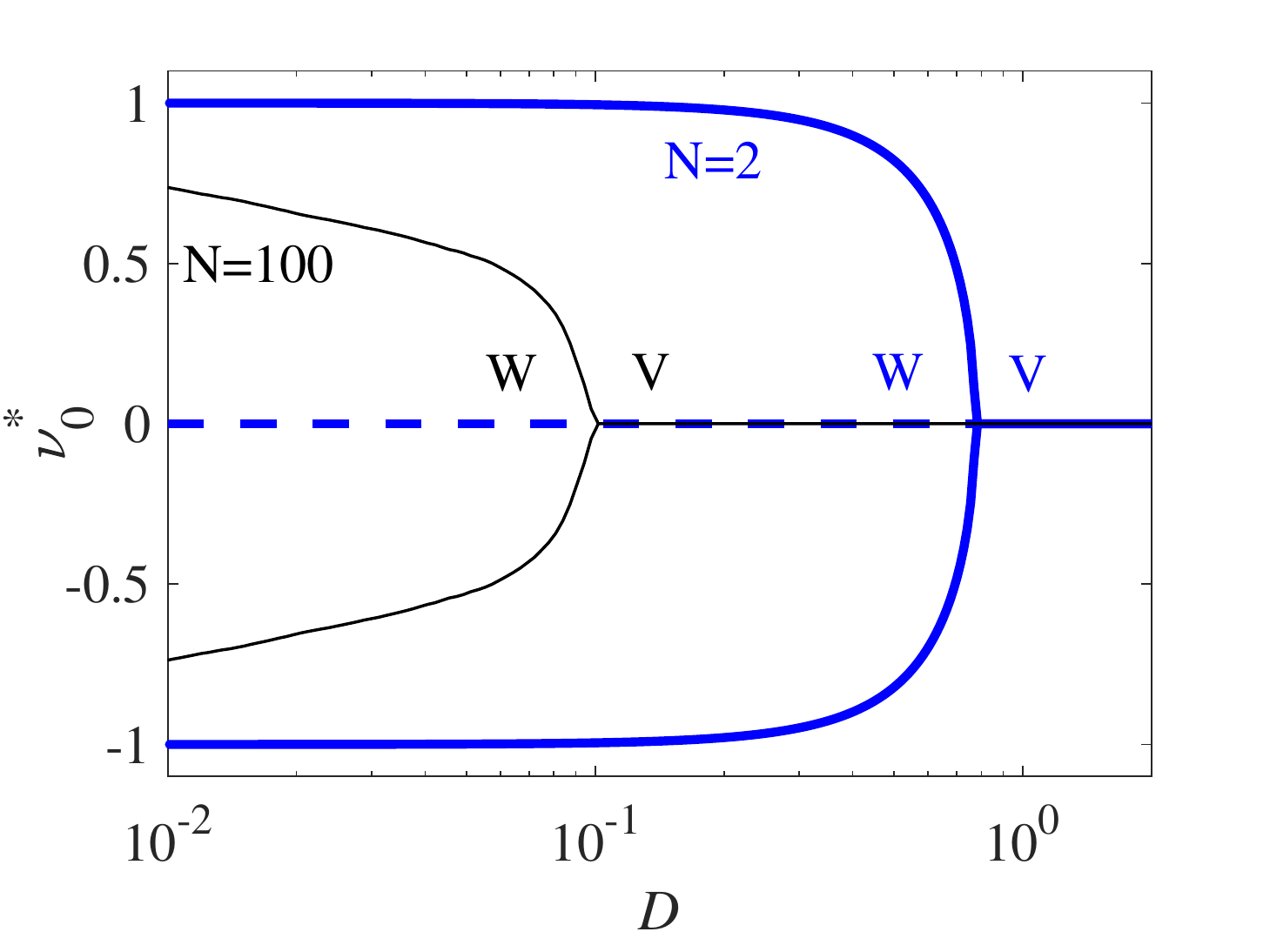}
		(b)\includegraphics[width=0.9\linewidth]{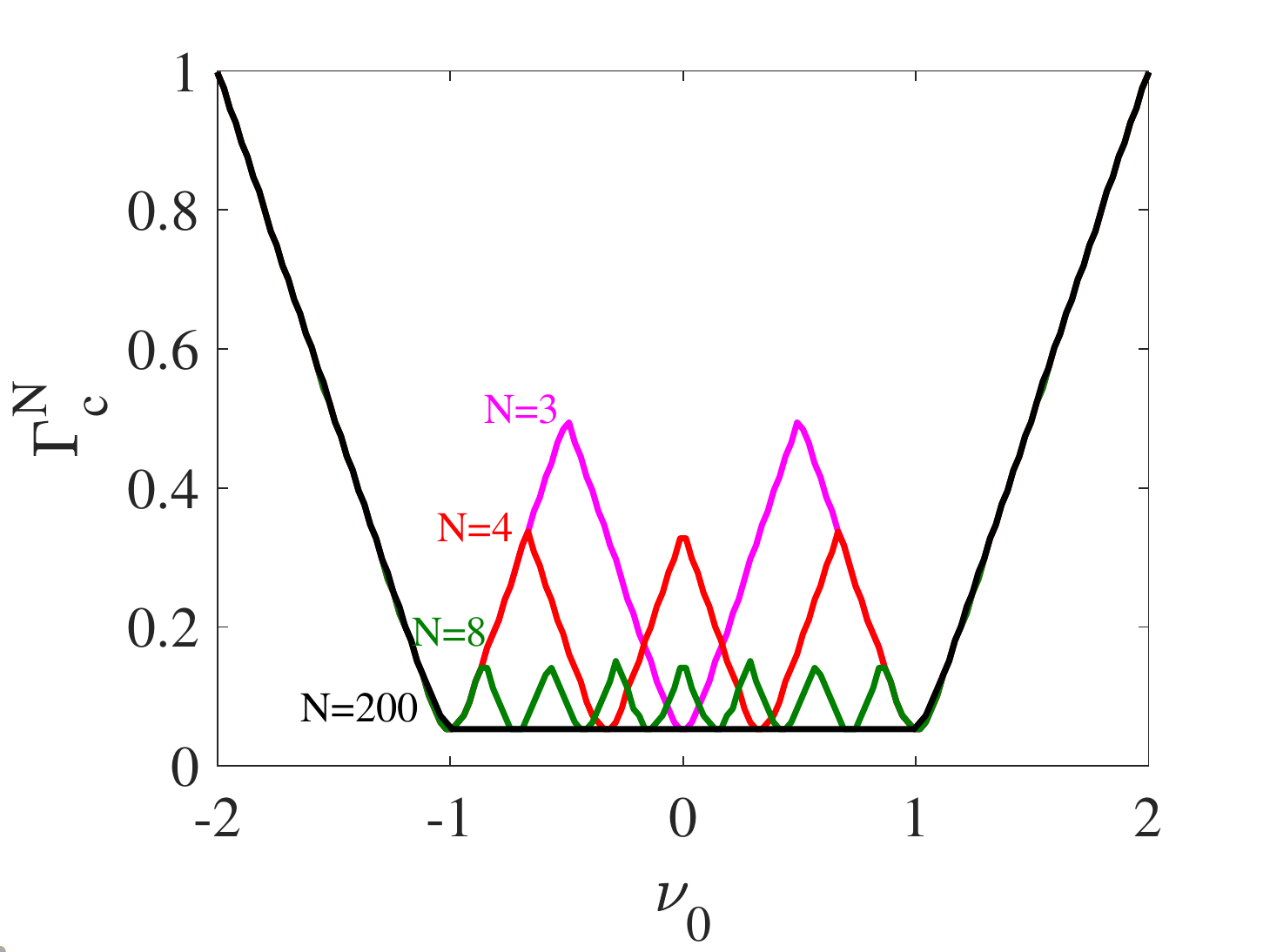}
		\caption{(a) Extremum points $\nu^*_0$ of the onset curve for $\Gamma^N_c$ as a function of $D$, computed for two oscillators using~\eqref{eq:twooscextremal} and numerically for $N=100$. The transition from 'W' to 'V' shape with the increase in the coupling strength is demonstrated. (b) Resonance tongue in the case of $D\to0$ for different number of oscillators, obtained using~\eqref{eq:ND0}. For both Figures, $\eta=1$.}
		\label{fig:diffn_tong}
	\end{figure}

	We start the discussion about discrete forms of resonance domains, with two coupled oscillators and specifically exploiting the analytical form \eqref{eq:threshold2Osc}. As has already been shown in Fig.~\ref{fig:osc2diffnu0}, two coupled oscillators are a special case due to the absence of the `U' shape region. Yet, the analysis is informative in the context of transition from `W' to `V' shape region (Fig.~\ref{fig:newtongue}(a)). The transition in the resonance tongue shape is in fact, similar to transition from double-- to single--well potential in the theory of phase separation (a.k.a. spinodal decomposition). Thus, to identify the extremum points of the tongue, we solve for $\partial (\Gamma_{c,2})^2/\partial\nu_0=0$ and find that three solutions
	\begin{equation}\label{eq:twooscextremal}
	\nu_0^{\ast}=0,\quad \pm\eta^{-1}\sqrt{\eta^4-D^2\Bra{\eta^2+(\mu-D)^2}}
	\end{equation}
	can coexist for:
	\begin{equation}
	1+\frac{(\mu-D)^2}{\eta^2}<\bra{\frac{\eta}{D}}^2.
	\end{equation}
	Specifically, in the limit $|\mu|\ll D$, we obtain $D<D_c=\eta\sqrt{(\sqrt{5}-1)/2} \simeq 0.786 \eta$, as marked in Fig.~\ref{fig:diffn_tong}(a). We also show that for $D\to\infty$, the transition from `W' to `V' shape persists with increasing in $N$ but shifts toward lower values of $D$, see Fig.~\ref{fig:theEffectOfnu0}.
	
	Next, we turn to the vanishing $D$ limit, where we have already indicated that the shape of the resonance domains strongly depends on discretization (Fig.~\ref{fig:diffn}). In contrast to the continuous limit, where the resonance is associated with the location of the hump, $x_0$, here in the discrete case the resonant location depends on the adjacent oscillator that has detuning closest to $\nu_0$. This leads to a jagged structure in $\Gamma_c$ given by
	\begin{equation}\label{eq:ND0}
	\Gamma^N_c=\sqrt{\mu^2+\bra{\min\Bra{\nu^j_L-\nu_0}}^2},
	\end{equation}
	where the minima correspond to values of $\nu_0$ equal to $\nu_L^j$. The number of oscillators that correlate with the number of extremum points is as following: For $N$ oscillators (and $|\nu_0|<|\eta|$) there are $2N-1$ extremum points, as shown in Fig.~\ref{fig:diffn_tong}(b). Consequently, for $N\gg1$ we get the `U' shape tongue as in the continuous case (Fig.~\ref{fig:newtongue}(a)).
	
	\section{Discussion}\label{sec:disc}
	
	In this paper, we studied the impact of spatial heterogeneity on the onset of resonant oscillations in damped oscillatory media subjected to parametric periodic forcing. {We used the universal amplitude equation for the Hopf bifurcation, which in the presence of periodic forcing takes the form of the  FCGL equation \eqref{eq:modelFCGL}.} For simplicity, we performed most of the analysis using linear spatial dependence of the detuning and also focused on a regime where only supercritical bifurcations are expected, i.e., setting $\beta=0$ in~\eqref{eq:FCGL}, whereas $\beta \neq 0$ is discussed elsewhere~\cite{ourPhysD}.
	
	We first studied how the the threshold of resonant oscillations, $\Gamma_c$, depends on the monotonic spatial heterogeneity (linear, convex or concave) of the detuning, $\nu$, and on the spatial coupling strength, $D$. We then addressed the case of discrete oscillator chains. In the case of a continuous oscillatory medium we showed that the classical `V' shape resonance tongue may also take a `W' form that resembles an ``inverse camel'' shape. Additionally, the `W' shape may become a `U' if the spatial coupling is weak. 	The implication of the classical `V' shape resonance in a monotonically heterogenous medium is localized oscillations, restricted to the location where the natural frequency is in resonance with the forcing. The new `W' and `U' shape resonances then imply partial delocalization of the oscillations. In the discrete oscillators case we studied the difference between even and an odd number of oscillators within the chain and showed that by increasing the number of oscillators (for a fixed length), we can recover the results of the continuous limit.

\begin{figure}[tp]
	\includegraphics[width=0.9\linewidth]{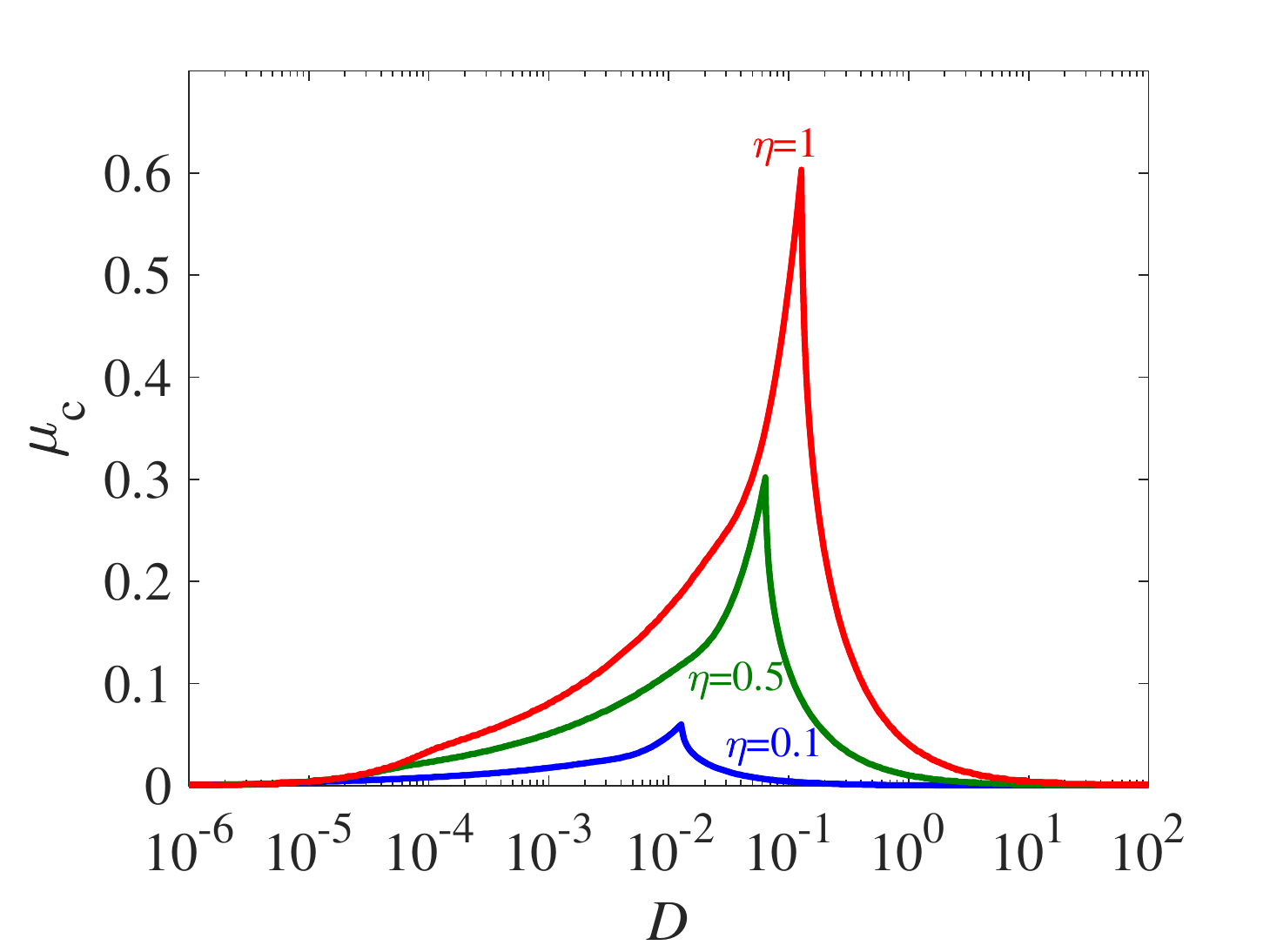}
	\caption{Oscillation thresholds as functions of the coupling strength $D$ for a continuous medium but without external parametric forcing, that is $\Gamma=\nu_0=0$ in~\eqref{eq:modelFCGL}. The thresholds have been computed using~\eqref{eq:lns} for three values of $\eta$.} \label{fig:unforced}
\end{figure}

{Reflecting back on sound detection in the auditory system, a `U' shape tongue implies similar detection sensitivity to a range of incoming frequencies, while the `W' shape tongue indicates location-dependent sensitivity along the cochlea or distinct levels of sensitivity for different frequencies. While we confined ourselves in this study to purely parametric forcing, there is a need for more realistic models of the cochlea that include combined (additive and parametric) forcing terms, from which amplitude equations can be derived and analyzed. The conditions that give rise to `W' and `U' shape resonances may then be related to physiological parameters and the implications to sound discrimination may possibly be studied within these lines. 
  
We used throughout this study Neumann boundary conditions, but other boundary conditions, such as Dirichlet, can be considered too. In Appendix A, we briefly discuss some differences between these two cases in terms of the shape of the resonance tongue and the spatial profile of the localized oscillations. We also focused on spatial heterogeneities that are introduced through system parameters, mostly the natural frequency of the oscillations, but the heterogeneity can also be introduced through the forcing amplitude. In Appendix B we address the case where the applied external forcing has a spatial Gaussian shape, which has recently been studied in the context of parametrically forced surafce waves (Faraday waves)~\cite{moriarty2011faraday,urra2019localized}.}

The applied methodology and the insights that have been obtained, can also be used to shed new light on the behaviors of `oscillation death' and `oscillation birth' in unforced media~\cite{kedma1985,aronson1990amplitude,koseska2013oscillation}. Setting $\Gamma=\nu_0=0$ in~\eqref{eq:modelFCGL} or in~\eqref{eq:discrete}; the oscillation threshold in the unforced system is the distance from the Hopf bifurcation, that is, from $\mu=0$. In Fig. ~\ref{fig:unforced}, we explicitly show that a qualitatively similar hump-type shape (compare with Fig.~\ref{fig:theEffectOfEta}) can also be obtained in the unforced case. The result is consistent with previous works that explored synchronization properties of oscillator chains but extends them to heterogeneous media, showing that moderate coupling strength indeed has destructive features even if the system is above the onset of the Hopf bifurcation.

{More broadly, we believe that the results and insights obtained here, can apply to other contexts of resonant oscillations that admit or are tunable to include spatial inhomogeneities. Such applications range from plant communities subjected to seasonal forcing~\cite{nathan2016linking,tzuk2019interplay}, through
mechanical resonators (NEMS and MEMS)~\cite{lifshitz2010nonlinear,jia2013parametrically,abrams2014nonlinear} to Kerr-type optical parametric oscillators~\cite{longhi1996stable,jang2019observation}.}

\begin{acknowledgments}
	This research was financially supported by the Ministry of Science and Technology of Israel (grant no. 3-14423) and the Israel Science Foundation under Grant No. 1053/17. Y.E. also acknowledges support by the Kreitman Fellowship.
\end{acknowledgments}

{	
\subsection*{Appendix A: The effect of boundary conditions}

\begin{figure}[tp]
	\includegraphics[width=0.9\linewidth]{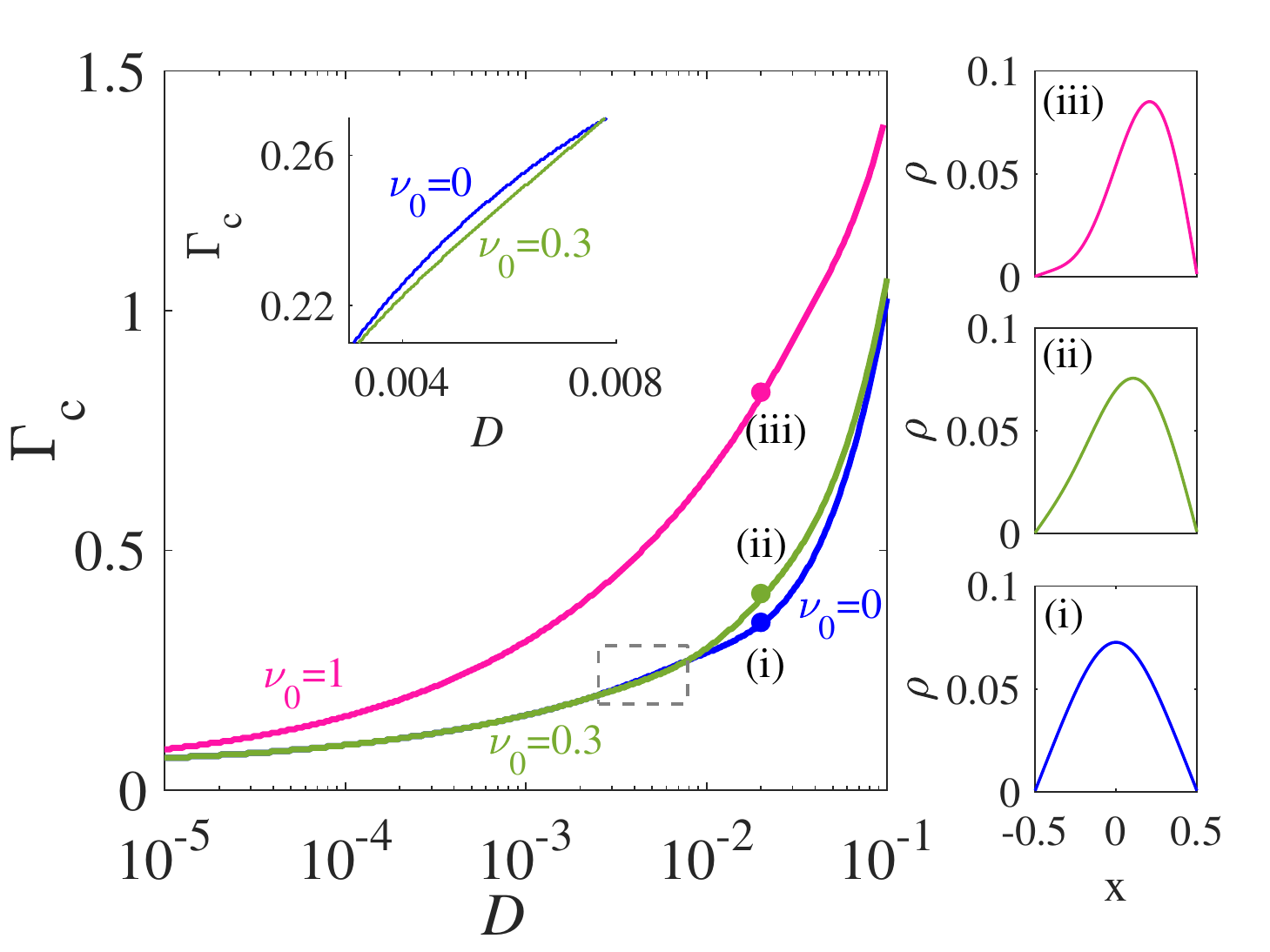}
	\caption{Resonance onset ($\Gamma_c$) as a function of spatial coupling strength ($D$) at several $\nu_0$ values under Dirichlet boundary conditions (by solving numerically~\eqref{eq:lns} with $\eta=1$). The inset shows a zoom-in of a region that is marked by the dashed rectangle while the profiles on the right, correspond to the respective 
solid circles on the $\Gamma_c$ curves at $D=0.02$.}\label{fig:diffnu0Diri}
\end{figure}
	
To demonstrate the effect of different boundary conditions (BC) we consider~\eqref{eq:lns} with the Dirichlet BC, $a(x=\pm 1/2)=0$. Unlike the non-monotonic behavior of $\Gamma_c$ with increased $D$ that was found with Neumann BC (see~Fig.~\ref{fig:theEffectOfnu0}), Direchlet BC give rise to a monotonic increase of $\Gamma_c$ with $D$, as shown in Fig.~\ref{fig:diffnu0Diri}. In addition, the size of the `W' shape region of the resonance tongue is narrowed down, as the inset in Fig.~\ref{fig:diffnu0Diri} indicates. This effect looks negligible, but it can be made stronger by increasing $\eta$, or by employing convex or concave heterogeneities (not shown here).

\begin{figure}[tp]
	\includegraphics[width=0.9\linewidth]{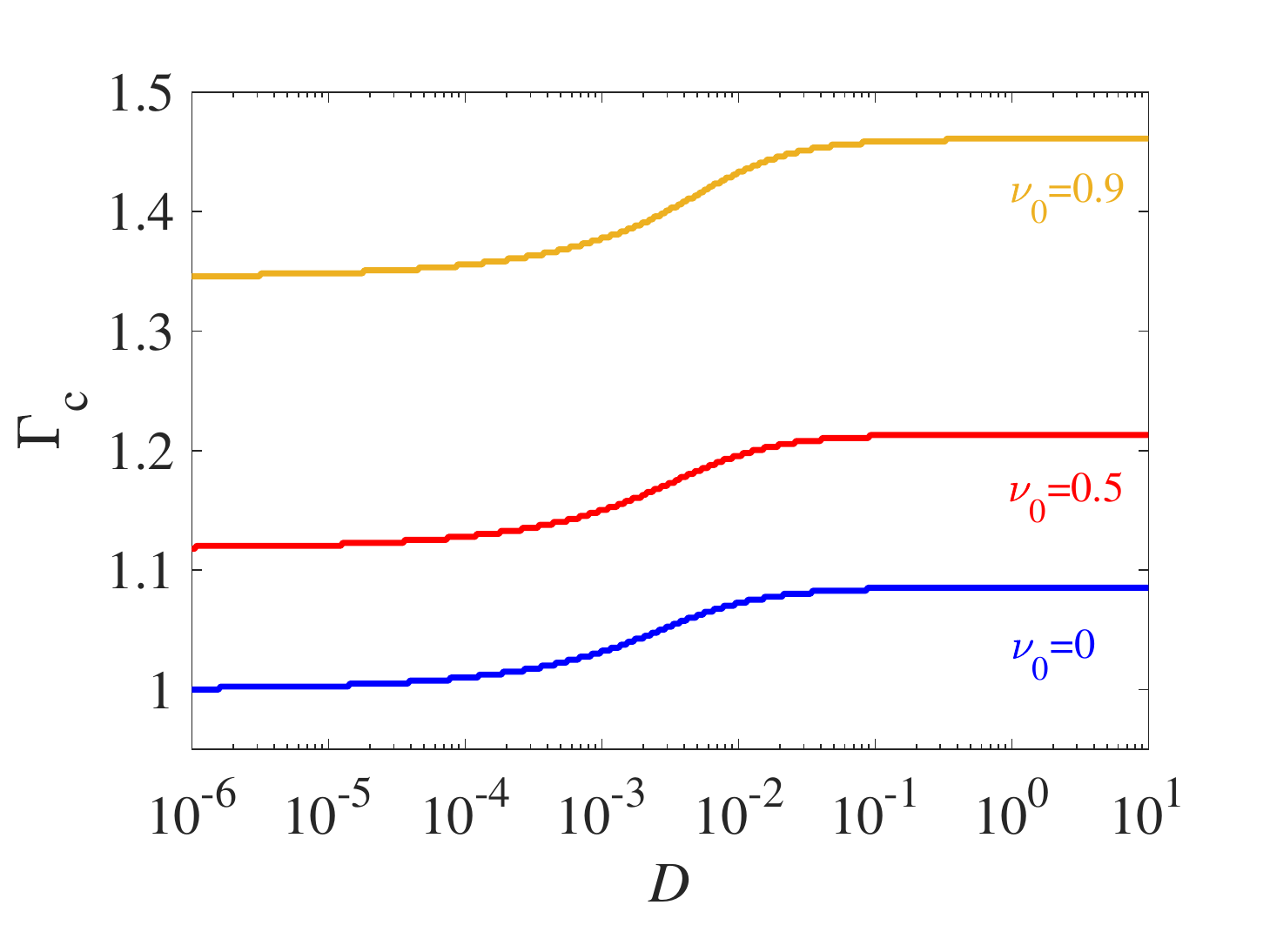}
	\caption{Resonance onset ($\Gamma_c$) as a function of spatial coupling strength ($D$) at several $\nu_0$ values for Gausian dependence in space of the applied forcing; the results were obtained numerically by solving~\eqref{eq:lns} with~\eqref{eq:forcing} and $\eta=0$.}
	\label{fig:onsetvsdfordiffnu0mux}
\end{figure}
	
\subsection*{Appendix B: Heterogeneity in the applied forcing amplitude}

Let us consider a spatially uniform detuning, but space dependent forcing amplitude. Specifically, let us assume a Gausian dependence:
\begin{equation}\label{eq:forcing}
\Gamma_x(x)=\Gamma g(x)=\Gamma e^{-x^2}.
\end{equation}
Applying the methodology used here, we first compute the resonance threshold for the limiting cases of weak ($D\to0$) and strong ($D\to \infty$) spatial coupling:
\begin{equation}\label{onsetGpofx}
\Gamma_c  = \left\{ {\begin{array}{cl}
	{\dfrac{\sqrt{\mu^2+\nu_0^2}}{\max[g(x)]},} & \text{for }D\to 0  \\ \\
	{\dfrac{\sqrt{\mu^2+\nu_0^2}}{\avr{g(x)}},} & \text{for }D\to\infty \\
	\end{array}}, \right.
\end{equation}
where $\avr{g(x)}$ is the spatial average of $g(x)$.

In both limits, the resonance onset, $\Gamma_c$, monotonically increases with $|\nu_0|$, that is, has the typical `V' shape of the resonance tongue. Complementary interpolation using a similar calculation to the one performed above, indicates that the `V' shape persists for any value of $D$, as shown in Fig.~\ref{fig:onsetvsdfordiffnu0mux}. A qualitatively similar behavior is obtained for linear spatial dependence of the distance from the Hopf bifurcation, $\mu(x)$ (not shown here).
}

\providecommand{\noopsort}[1]{}\providecommand{\singleletter}[1]{#1}%

\end{document}